\documentstyle[sprocl,epsfig]{article}

\def\3he{{$^3${\rm He}}}

\def\ie{{\it i.e.,\ }}

\def\slD{\raise.15ex\hbox{$/$}\kern-.53em\hbox{$D$}}
\def\dsl{\raise.15ex\hbox{$/$}\kern-.57em\hbox{$\Delta$}}
\def\slp{{\raise.15ex\hbox{$/$}\kern-.57em\hbox{$\partial$}}}
\def\nsl{\raise.15ex\hbox{$/$}\kern-.57em\hbox{$\nabla$}}
\def\sla{\raise.15ex\hbox{$/$}\kern-.57em\hbox{$\rightarrow$}}
\def\slla{\raise.15ex\hbox{$/$}\kern-.57em\hbox{$\lambda$}}
\def\slb{\raise.15ex\hbox{$/$}\kern-.57em\hbox{$b$}}
\def\lnp{\raise.15ex\hbox{$/$}\kern-.57em\hbox{$p$}}
\def\lnk{\raise.15ex\hbox{$/$}\kern-.57em\hbox{$k$}}
\def\lnK{\raise.15ex\hbox{$/$}\kern-.57em\hbox{$K$}}
\def\lnq{\raise.15ex\hbox{$/$}\kern-.57em\hbox{$q$}}

\def\be{{\beta}}

\def\cO{{\cal O}}

\def\pmb#1{\setbox0=\hbox{$#1$}%
\kern-.025em\copy0\kern-\wd0
\kern.05em\copy0\kern-\wd0
\kern-.025em\raise.0433em\box0 }

\def\ee{{e^+e^-}}

\def\q2{{Q^2}}
\def\gtwid{\raise.3ex\hbox{$>$\kern-.75em\lower1ex\hbox{$\sim$}}}
\def\ltwid{\raise.3ex\hbox{$<$\kern-.75em\lower1ex\hbox{$\sim$}}}
\def\12{{1\over2}}
\def\part{\partial}

\def\low#1{\lower.5ex\hbox{${}_#1$}}

\def\psl{\raise.15ex\hbox{$/$}\kern-.57em\hbox{$\partial$}}
\def\partt{\raise.15ex\hbox{$\widetilde$}{\kern-.37em\hbox{$\partial$}}}

\def\topppageno1{\global\footline={\hfil}\global\headline
={\ifnum\pageno<\firstpageno{\hfil}\else{\hss\twelverm --\ \folio
\ --\hss}\fi}}
 
\def\toppageno2{\global\footline={\hfil}\global\headline
={\ifnum\pageno<\firstpageno{\hfil}\else{\rightline{\hfill\hfill
\twelverm \ \folio
\ \hss}}\fi}}

\def\prd#1{Phys.\ Rev.\ {\bf D#1}}

\def\ie{{\it i.e.},\ }

\def\et{{\it et al.}}

\def\nsection#1 #2{\leftline{\rlap{#1}\indent\relax #2}}

\def\prd#1{Phys.\ Rev.\ {\bf D#1}}

\def\seillac{Nucl.\ Phys.\ {\bf B} (Proc.\ Suppl.) {\bf 4} (1988)}

\def\amsterdam{Nucl.\ Phys.\ {\bf B} (Proc.\ Suppl.) {\bf 30} (1993)}
\def\dallas{Nucl.\ Phys.\ {\bf B} (Proc.\ Suppl.) {\bf 34} (1994)}
\def\bielefeld{Nucl.\ Phys.\ {\bf B} (Proc.\ Suppl.) {\bf 42} (1995)}
\def\melbourne{Nucl.\ Phys.\ {\bf B} (Proc.\ Suppl.) {\bf 47} (1996)}

\def\stlouis{Nucl.\ Phys.\ {\bf B} (Proc.\ Suppl.) {\bf 53} (1997)}

\def\tsukubanew{talk presented at the International Workshop
on Lattice QCD, Tsukuba, Japan, March 10-15, 1997,
to be published}

\def\edinburgh{talk presented at the International Symposium,
{\it Lattice '97}, Edinburgh, UK, July 22--26, 1997, to be published
in Nucl.\ Phys.\ {\bf B} (Proc.\ Suppl.)}

\def\Journal#1#2#3#4{{#1} {\bf #2}, #3 (#4)}

\def\NPB{{\em Nucl.\ Phys.}\ B}
\def\PLB{{\em Phys.\ Lett.}\ B}

\def\PRD{{\em Phys.\ Rev.}\ D}
\def\ZPC{{\em Z.\ Phys.}\ C}

\def\be{\begin{equation}}
\def\ee{\end{equation}}
\def\bea{\begin{eqnarray}}
\def\eea{\end{eqnarray}}
\newcommand{\fB}{$f_B$}
\newcommand{\fBs}{$f_{B_s}$}
\newcommand{\fD}{$f_D$}
\newcommand{\fDs}{$f_{D_s}$}
\newcommand{\fBsofB}{$f_{B_s}/f_B$}
\newcommand{\fDsofD}{$f_{D_s}/f_D$}

\newcommand{\fBsofDs}{$f_{B_s}/f_{D_s}$}
\newcommand{\fBofDs}{$f_B/f_{D_s}$}

\begin{document}

\title{LATTICE CALCULATIONS OF DECAY CONSTANTS \footnote{review 
presented at the 
{\it Seventh International Symposium on Heavy Flavor Physics}, 
Santa Barbara, July 7-11, 1997}
}

\author{ C.\ BERNARD
}

\address{Department of Physics, Washington University, St.\ Louis, MO
63130, USA}


\maketitle\abstracts{
Lattice attempts to compute the leptonic decay constants of
heavy-light pseudoscalar mesons are described.  
I give a short historical overview  of such attempts and then
discuss some current calculations. I focus on three of the most important
sources of systematic error:  
the extrapolation to the continuum,
the chiral extrapolation in light quark
mass, and the effects of quenching.
I briefly discuss the ``bag parameters''
$B_B$ and $B_{B_s}$, and then conclude with my expectations of the
precision in decay constants and bag parameters
that will be possible in the next few years.}
  
\section{Historical Overview}

Over the past ten years, many groups have
attempted to calculate \fB, \fBs, \fD\ and \fDs\ from lattice QCD.
While the predictions for \fD\ and \fDs\ have been quite stable,
those for \fB\ and \fBs\ have had a rather checkered history.
It may be useful therefore to give a brief review of this
history and to point out some of the problems that early computations
encountered.  I've divided my history into three periods:
``ancient history,''  the ``middle ages,'' and the ``modern era.''
Let me caution at the outset that my divisions are somewhat arbitrary,
and my placing of various papers into one period or another
quite subjective.  Furthermore, I do not pretend to have included
all the papers of this subject; I have tried only to include what
I consider ``representative'' works.
For more comprehensive and detailed reviews, see Refs.~[1,2].

\vspace{-.3cm}
\subsection{Ancient History}\label{subsec:ancient}
\vspace{-.15cm}

While much that was useful was learned, the results for \fB\ and
\fBs\ that came from several early computations may now be ignored.
First among these was a computation I was involved in: Ref.~[3] (``BDHS'').
We obtained a very low value of
$f_B \sim 100$ MeV.  The problem came from the
treatment of heavy quarks on the lattice.  While the
results for \fD\ and \fDs\ were reasonable, the extrapolation from
the $D$ to the $B$ was strongly skewed by lattice artifacts,
which arise whenever the heavy quark mass ($m_Q$) is comparable to the inverse
lattice spacing ($a^{-1}$).

The static approximation on the lattice, in which $m_Q\to\infty$,
is an approach that was
introduced \cite{eichten} to avoid the problems due to
$m_Qa\sim 1$.  However early computations within the static
approximation \cite{elc91,pcw91,ape94} obtained very large
answers: $f_B\sim 300$ MeV.  I believe it is now generally
accepted that the main problem here was contamination by
excited states.  The static approach is inherently very noisy,
and without sophisticated techniques \cite{variational} that allow one to 
extract the signal at short times, one may easily be fooled
by ``false plateaus'' into believing that excited state contamination
is under control.  In addition, it is now known that the result
for $f_B$ at the physical $B$ meson is $\sim 20\%$ lower than
at infinite $b$ quark mass, and that the extrapolation to the continuum
also tends to lower $f_B$.

\vspace{-.2cm}
\subsection{Middle Ages}\label{subsec:middle}
\vspace{-0.15cm}

In this middle period I place papers which have no glaring problems,
but which, either because of limited resources or limited focus,
have in general rather crude systematic error estimates.  This is not
to say that systematic errors are ignored or treated cavalierly.
Indeed, some of these errors are studied in great detail.  However,
even if one ignores the errors due to ``quenching'' (the neglect
of virtual quark loops, \ie of sea quarks),  none of these computations
is able to include a complete study of all other significant sources
of error. In Table \ref{tab:middle}, I list the
papers  that I have placed in this period, together with 
a list of relevant systematic issues. 

A frequent concern about the papers in Table \ref{tab:middle}
is the lack of a reasonable
way to estimate the errors coming from the
extrapolation to the continuum.  Often only one lattice spacing is available; 
when there are two, large statistical errors usually hide any 
systematic difference.  Only two of the computations \cite{fnal95,pcw94}
consider  a wide range of lattice spacings, which allows some control
over the extrapolation.  
%
Note that, even for
computations \cite{nrqcd,nrqcd-japan}
 which treat the heavy quark in the
effective nonrelativistic QCD (NRQCD) framework,
\cite{lepage}
an error estimate of discretization effects 
is needed because the light quark is still treated with a standard
relativistic action. These discretization errors are not included in
the truncation errors of the NQRCD action.

A second issue for several of the papers in Table~\ref{tab:middle}
are the artifacts which occur for $m_Qa\sim1$ in a conventional,
propagating heavy quark
approach, and which skewed the results of Ref.~[3].
In Ref.~[14],
it is shown how such artifacts can be corrected
for all $m_Qa$, 
order by order in $ap$ (where $p$ is a typical 3-momentum) and
$\alpha_s(q^*)$ (where $q^*$ is a scale of the order of $a^{-1}$).
The most important of these correction is
the so-called ``EKM'' factor. Without this factor, the 
the heavy-light decay constants become inconsistent with the
static-light decay constants at fixed $a$ once $m_Qa\,\gtwid\, 0.5$.~\cite{bls}
Refs.~[16,10]
do not make this correction.  However, 
the rather large statistical errors mask somewhat the inconsistency,
and the final results, which more or less average between
the propagating and static quark results, are not unreasonable.

\begin{table}[t]
\renewcommand{\arraystretch}{1.3}
\caption{Examples of decay constant calculations from the ``middle
ages:'' no major problems, but incomplete study of systematic
errors.  All of the calculations were performed in the quenched
approximation, and none of them tried to estimate quenching
errors.  The other concerns listed are: (A) no continuum extrapolation,
(B) inconsistent  propagating and static quark normalization,
(C) small physical volumes, (D) static approximation only, 
(E) $\cO(1/M^2)$ operators neglected, (F) complete 1-loop perturbative
correction not included,
(G)~lattice spacing effects on light quarks not estimated.
\label{tab:middle}}
\vspace{0.4cm}
\begin{center}
\begin{tabular}{|c|c|c|}
\hline
 & & \\
\bf Group &\bf $f_B$ (MeV)&\bf Concerns
\\ \hline\hline
ELC \cite{elc92}& $205(40)$& A, B
\\ \hline
UKQCD \cite{ukqcd}& $160({}^{+6}_{-6})({}^{+59}_{-19})$& A \\
& $176({}^{+25}_{-24})({}^{+33}_{-15})$& 
\\ \hline
BLS \cite{bls}& $205(40)$& A
\\ \hline
PCW \cite{pcw94}& $180(50)$& B, C
\\ \hline
FNAL \cite{fnal95}& $188(23)(15)({}^{+26}_{-0})(14)$& D
\\ \hline
APE \cite{ape97}& $180(32)$& A
\\ \hline
NRQCD \cite{nrqcd}& $183(32)(28)(16)$&  E, G
\\ \hline
Hiroshima \cite{nrqcd-japan}& $184(7)(5)(37)(37)$&  F, G
\\ \hline
\end{tabular}
\end{center}
\end{table}

\vspace{-.2cm}
\subsection{Modern Era}
\vspace{-.15cm}

I place in the ``modern era'' those computations which 
make a serious attempt to quantify all the sources of systematic
error, at least within quenched approximation.  In addition, one of
these calculations \cite{milc-b20} tries to estimate the error
due to quenching by comparing some quenched results with those on
comparable lattice that include virtual quark effects.  
However the resulting estimates remain
rather crude at the moment --- see below.  The three on-going
computations which I place in the modern category are listed in
Table~\ref{tab:modern}, along with some representative results and
my attempt to average them.
It must be emphasized that the results from all three groups are
still preliminary.  I note that the recent computation 
by the APE group \cite{ape97}
could also be considered ``modern'' by many criteria.  However, I 
have placed it in the previous section because, with only two
values of the lattice spacing, it is difficult to draw any conclusions
about the size of the systematic error of the continuum extrapolation.

\begin{table}[t]
\renewcommand{\arraystretch}{1.3}
\caption{Decay constant calculations from the ``modern
era.'' 
All of the calculations are preliminary.  
The first error in all cases is statistical; the second represents
systematic errors within the quenched approximation.
The third error from MILC is an estimate of the quenching
effect; the other groups do not consider quenching errors
at this time. 
The systematic
errors for JLQCD do not yet include the effects from the
continuum extrapolation, estimated at ``a few percent,''
and the errors from weak coupling perturbation theory,
estimated as 5\%. ``BraveWA'' are my brave
(or foolhardy!) attempts at world
averages, which ignore certain 
systematic inconsistencies
in the data (see text), and assume that the quenched approximation
results
can be {\em corrected} by the MILC quenching error when this
has a definite sign. The errors
in world averages are combined total errors; a more conservative
approach would result in errors larger by a factor
of about $1.5$.\label{tab:modern}}

\vspace{0.4cm}
\begin{center}
\begin{tabular}{|c|c|c|c}
\hline
 & & & \\
\bf Group &\bf $f_B$ (MeV)&\bf \fBsofB & \bf \fDs\ (MeV)
\\ \hline\hline
MILC \cite{milc-b20}& $153(10)({}^{+36}_{-13})({}^{+13}_{-0})$& 
$1.10(2)({}^{+5}_{-3})({}^{+3}_{-2})$& 
$199(8)({}^{+40}_{-10})({}^{+10}_{-0})$
\\ \hline
JLQCD \cite{jlqcd}& $163(12)({}^{+13}_{-16})$& 
 $-$&
 $213(11)({}^{+12}_{-18})$
\\ \hline
FNAL \cite{fnal97}& $166(10)(28)$& 
$1.17(4)(3)$& 
$215(7)(30)$
\\ \hline
BraveWA& $175(30)$& 
$1.14(5)$& 
$221(25)$
\\ \hline
\end{tabular}
\end{center}
\end{table}

\section{Systematic Errors}

The sources of the three largest systematic errors affecting the
calculation of the MILC collaboration~\cite{milc-b20} are the 
extrapolation to the
continuum, the ``chiral" extrapolation (extrapolation from the
lattice light --- $u$ and $d$ --- quark masses to the physical light quark
masses), and the quenched approximation.  Here, I will briefly discuss each
of these systematic errors.  (For a more comprehensive discussion, 
see Ref.~[19].)
As will become
clear below, the dominant sources of error in the other two 
``modern'' calculations  may be different.

\vspace{-.2cm}
\subsection{Continuum Extrapolation}
\vspace{-0.15cm}

\begin{figure}[b!] 
\vskip -1.5cm
\centerline{\epsfig{file=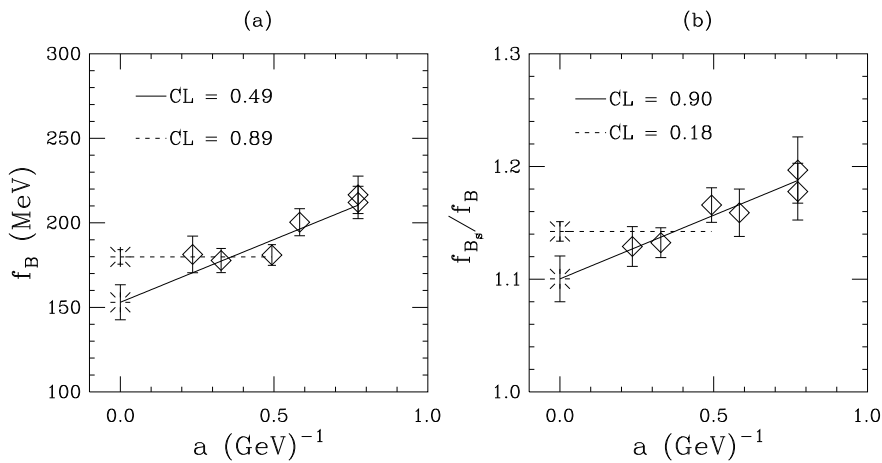,height=3.1in,width=3.1in}}
\caption{(a) Preliminary quenched results from the MILC collaboration
for \fB\ as a function of lattice spacing
$a$.
The errors bars reflect only statistical errors and the errors from
isolating the state of interest.
The diamonds are the lattice data points, and the bursts are
extrapolated values.
The linear fit to all the diamonds (solid line) gives the central value; 
a constant fit to the three smallest lattice spacings (dotted line)
gives one way to estimate the systematic error of the extrapolation.
``CL'' denotes confidence level.  (b) Same as (a), but for the
ratio \fBsofB\ {\it vs.} $a$.
\label{fig:fb}}
\end{figure}

With the Wilson fermions used in the MILC computation, the leading
errors as $a\to0$ are $\cO(a)$, although the coefficient of $a$, as
well as the size of higher order terms, are
of course unknown  {\it a priori}.  
In Fig.~\ref{fig:fb}, I show this
extrapolation for \fB\ and the ratio \fBsofB.  Assuming linearity
in $a$ over the full range of lattice spacings studied gives
the central values.  Another plausible assumption is that for the
three smallest spacings (corresponding to $\beta=6.0$, $6.3$, and $6.52$),  
terms of $\cO(a)$ and higher are already quite small.  This leads
to a constant fit to these three points, and we take the difference
between the two fits as one measure of the systematic error.
Other measures of the error
are obtained by comparing the linear fit to all points
to a {\em linear} fit to the three points with smallest lattice spacing, 
and by
varying or omitting the EKM corrections and repeating the linear fit
to all points.  (For more details see Ref.~[19].)

For all the decay constants as well as almost all ratios of
decay constants (the exception is \fBofDs), the difference  of
linear and constant fits, as in Fig.~\ref{fig:fb}, gives the largest
error estimate and is therefore taken to be the systematic error
of the continuum extrapolation. For the decay constants themselves,
this error is rather large ($\sim\!12$--$27\%$), reflecting the fact that the
slope in $a$ is steep.  The ratios of decay constants
are  much better behaved with
an error of $\sim\!4$--$5\%$.  These errors are in general the largest
of all the systematic errors, for both decay constants and ratios.

\begin{figure}[b!] 
\vskip -2.0cm
\centerline{\epsfig{file=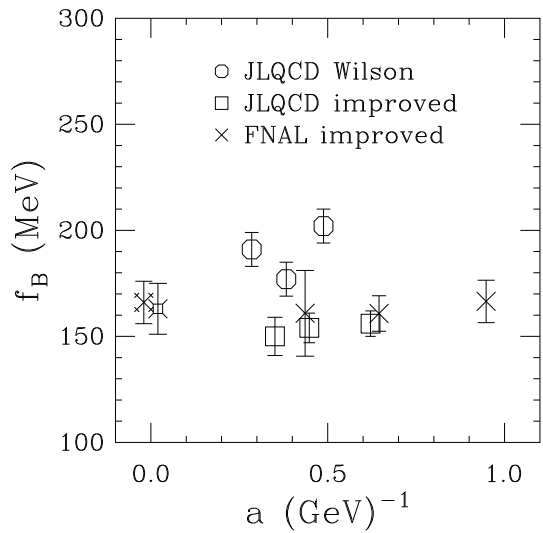,height=3.5in,width=3.5in}}
\caption{Preliminary quenched results from the JLQCD and FNAL collaborations
for \fB\ as a function of lattice spacing
$a$.  The errors shown are statistical only.
For JLQCD, results with both Wilson (octagons) and the improved
action (squares) are shown; FNAL (crosses) uses improved action
exclusively.  Final extrapolated values with statistical
errors are also shown: ``fancy square'' for JLQCD and
``fancy cross'' for FNAL.
\label{fig:fb_others}}
\end{figure}

The continuum extrapolation errors can be reduced by ``improving'' the
action  --- removing discretization errors order by order in $a$
by adding additional operators to the action (and correcting the
axial current).  New data with improved actions
has appeared recently.~\cite{jlqcd,fnal97}
The preliminary
results  are shown in Fig.~\ref{fig:fb_others}.

FNAL implements the full EKM program \cite{ekm}  at $\cO(a)$ and up
to tadpole improved tree-level in $\alpha_s$.  
JLQCD improves the action with ``clover'' (Sheik\-hole\-slami-Wohlert)
fermions,\cite{clover} and adjusts the normalization and
shifts from pole to kinetic mass {\it \`a la} EKM.~\cite{ekm}
As I understand it, 
the main difference from the full EKM program at this order 
is in the corrections to the axial current, but these
corrections are expected \cite{onogi} to have little effect
on the final results.  

In practice, the apparent excellent agreement
between the improved results of
JLQCD and FNAL seen in Fig.~\ref{fig:fb_others} is somewhat
misleading, since JLQCD uses $m_\rho$ to set the scale, and
FNAL, $f_\pi$.  Indeed, when JLQCD uses $f_\pi$ to set the scale,
they see steeper $a$ dependence.~\cite{onogi}
Some time will thus be needed until the dust settles and the
situation is clearer.
However it seems quite obvious from Figs.~\ref{fig:fb_others} and \ref{fig:fb}
 that the
extrapolation to the continuum will be much better controlled in
the improved case than
in the Wilson case.  One will be able to reduce the error still further
by demanding equality between the results of the two methods.

\vspace{-.2cm}
\subsection{Chiral Extrapolation}
\vspace{-0.15cm}
  
The lattice computations are typically
performed with light quark masses ($m_q$) in the range
$m_s/3\, \ltwid\, m_q\, \ltwid\, 2m_s$.  This is because using physical
$m_u$, $m_d$ would (a) require too much computer time,
(b) require too large a lattice, and (c) introduce spurious
quenching effects.~\cite{qchpt}
The ``chiral extrapolation'' is then the extrapolation in 
$m_q$ to physical
$m_{u,d}$.  One must extrapolate not only the heavy-light decay constants
and masses, but also, in general, one or more 
experimentally known light-light quantities
in order to set the scale of the lattice spacing and determine the
correct lattice light quark mass.
Typical light-light quantities needed are $m_\pi$ and either
$f_\pi$ or $m_\rho$. 

The chiral extrapolation is a  significant source of error, with the majority
of the error
coming from the light-light quantities.  This is because the light-light
quantities are more non-linear in quark mass.
For example, although lowest order chiral perturbation theory
predicts that $m^2_\pi$ is a linear function of quark mass, one sees,
at the current statistical level, small but
significant deviations from linearity that are not well understood.

The deviations from linearity
could be due to unphysical effects such as
the finite lattice spacing or the
residual contamination by excited states.  Even the more
``physical'' cause (chiral logs or higher order
analytic terms in chiral perturbation theory) are a source of
spurious effects
because quenched chiral logs are in general different from those in the
full theory.~\cite{qchpt} 

For the above reasons, the MILC collaboration 
presently fits quantities like $m^2_\pi$ to a linear
form, despite the poor confidence levels.  The systematic error
is estimated by repeating the analysis with quadratic
fits.  The latter are constrained fits, since unfortunately
the computation has been done with only three light quark masses. 
The systematic thus determined is $\le\!10\%$ for decay constants
on all quenched data sets
used to extrapolate to the continuum; usually it is $\ltwid 5\%$.  After
extrapolation to the continuum, the error is larger: $7\%$ to $15\%$.

I emphasize that our reasons for choosing linear chiral fits for the
central values are somewhat subjective, and it is possible that we
will switch to quadratic fits in the final version of the work.
To help us make the choice, we are studying a large sample of lattices
at $\beta=5.7$, with large volumes up to $24^3$.  On this sample we
have six light quark masses and have mesons with nondegenerate  as
well as degenerate quarks. Should the switch be made, it would
raise the central value for $f_B$ by 23 MeV, that for $f_{D_s}$
by 14 MeV,  and other decay constants by comparable amounts.
 The systematic
error within the quenched approximation would then become much more symmetric,
with the continuum extrapolation the dominant positive error and the
chiral extrapolation the dominant negative one.

It is worth noting here that the MILC and JLQCD are quoting rather different
(by $\sim\!15\%$) lattice scales from $f_\pi$
at the one coupling ($\beta=6.3$) they have in common.  
Although both groups use linear chiral fits, the chiral extrapolation
may explain  at least part of the discrepancy, since the masses of MILC's light quarks 
extend to  higher masses than those of JLQCD, and quadratic terms are therefore
more likely to be important for MILC.
Another possible cause is finite size effects (JLQCD's lattice
at $\beta=6.3$ has size $32^3$; MILC's, $24^3$).  However, the MILC study (mentioned above)
of various volumes at $\beta=5.7$ shows no significant finite size effects
at the relevant physical volumes, and appears to give an upper bound or such effects
that is smaller than what is needed to explain the discrepancy.
The groups are in the process of comparing raw numbers and normalizations to try to settle
this issue.  

\vspace{-.2cm}
\subsection{Quenching}
\vspace{-0.15cm}

The quenched approximation has one great advantage: it saves an enormous
amount of computer time.  However, it is not a true approximation,
since there is no perturbative expansion of the full theory
for which the quenched approximation is the first term.  Thus one
should think of it only as a model, and it is imperative to estimate
its errors, and ultimately to move beyond it.

To this end MILC has repeated the computations on lattices with virtual
quark loops included.  I emphasize however that such computations
are not yet ``full QCD.''  This is because (1) the virtual quark mass
is fixed and not extrapolated to physical up or down mass 
(2) the virtual quark data is not yet
good enough to extrapolate to $a=0$, and (3) there are two light flavors,
not three.  Thus the virtual quark simulations are used at this point only
for systematic error estimation, and the error estimate so obtained
must be considered rather crude. 

\begin{figure}[b!] 
\vskip -2.0cm
\centerline{\epsfig{file=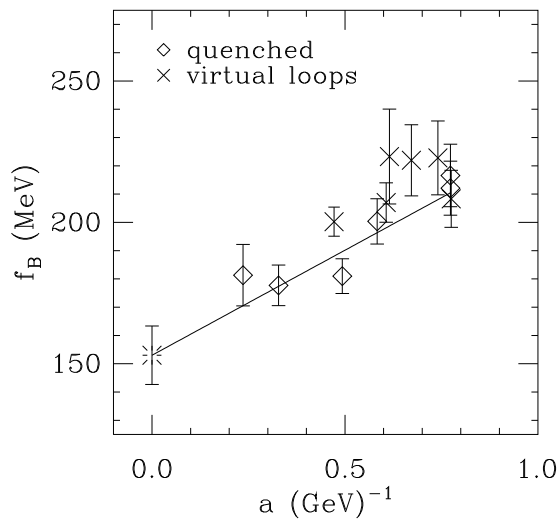,height=3.5in,width=3.5in}}
\caption{Preliminary MILC results for \fB\ as a function of lattice spacing
used for estimating the effects of quenching.  The diamonds
are the same as in Fig.~1(a).
The crosses (virtual quark loops included)
are not extrapolated to the continuum.  \label{fig:fb_dynamical}}
\end{figure}

Figure~\ref{fig:fb_dynamical} shows one way the quenching error is estimated.
We compare the
smallest-$a$ virtual quark simulation (the cross at
$a=0.47\ ({\rm GeV})^{-1}$) with the quenched simulations, interpolated
to the same value of $a$.  
See Ref.~[19]
for more details.

\section{Brief remarks on the B parameters}

Recent lattice computations \cite{soni,recentBB} 
of $B_{B_d}$ with propagating heavy quarks are consistent with
older work.~\cite{bdhs,oldBB,elc92}  They all give $B_{B_d}(5 {\rm GeV}) \approx
0.9$, with errors (within the quenched approximation)
of about 10\%.  However, the  results with static heavy quarks \cite{ukqcd96,kentucky,gimenez} 
vary widely, from about $0.5$ to $1.0$.  
The variation appears to be due in large part to the large uncertainties in
the perturbative corrections, in the matching both of lattice to continuum
and of static effective theory to full theory.  In addition, the
poor signal-to-noise properties of the static theory also may lead to some
difficulties.  (For compilations of both
propagating heavy-light and static-light results, see 
Refs.~[29,30].)

MILC has recently begun a computation of static-light  B
parameters.
Our very preliminary results give
$B_{B_d}(5 {\rm GeV}) \approx 0.9$.  This is consistent with the
results of the Kentucky group,\cite{kentucky} which is not surprising
since the perturbative corrections are treated in the same way.

None of the existing calculations (propagating or static)
is fully ``modern'' in the sense of
Section 1, since the extrapolation to the continuum has not yet been
studied carefully.  However, the indications are that $B_{B_d}$ has
only mild dependence on the lattice spacing.~\cite{soni}  If I also assume that
the Kentucky approach to the static-light perturbative corrections is
the most ``reasonable'' since it gives results that agree with the 
propagating heavy-light results, I arrive at
$B_{B_d}(5{\rm GeV}) = 0.86(3)(10)(8)$ 
for my (prejudiced) world average.  Here the first error is statistical,
the second summarizes systematics within the quenched approximation,
and the third is my {\it guess} of the quenching error. 
The renormalization group invariant, $\hat B_{B_d}$ at next to leading order
is then
$\hat B^{nlo}_{B_d} = 1.37(22)$.
Combining
this with the ``brave world average'' (Table~\ref{tab:modern})
 for \fB\ gives
\begin{equation}
\xi_d \equiv f_B\sqrt{\hat B^{nlo}_{B_d}} = 205(39)\ {\rm MeV}.
\end{equation}
The ratio $B_{B_s}/B_{B_d}$ is considerably more stable.  All groups
find it to be very close to 1.   Including the preliminary MILC results
(on lattices with and without virtual quarks),
I find a world average $ B_{B_s}/B_{B_d} = 1.00(1)(2)(2)$.
Combining this with the ``brave world average'' 
 for \fBsofB\ gives
\begin{equation}
\xi^2_{sd}\equiv  {f^2_{B_s} B_{B_s}\over f_B^2B_{B_d}} = 1.30(12)
\end{equation}
I emphasize that the errors in Eqs.~(1) and (2) are based on the optimistic
assumptions that go into the ``brave world averages.'' A more conservative 
approach would result in errors larger by a factor of about 1.5.

I note that it is also possible to calculate $\xi^2_{sd}$ directly,\cite{blum}
without calculating decay constants and B parameters separately.
We are getting (still preliminary) $\xi^2_{sd} = 1.68(10){}^{+43}_{-38}$.
Given the large errors, this is consistent with Eq.\ (2).
The direct method has, however, several advantages, and ultimately
may be competitive with or even superior to the separate computations.

\section{The ``Post-Modern Era''}

I expect a significant improvement over the next few years in
the lattice evaluation of decay constants.  Within the quenched
approximation, there should be good control of all the systematic
errors.  In particular, 
improved actions (and possibly smaller lattice spacings) will allow
for a considerable reduction of the
continuum extrapolation error. Further, the chiral
extrapolations should be better controlled by using smaller quark
masses on physically larger lattices. (JLQCD has already been able to
move in this direction.)  Within two or three years, the conservative
total error within the quenched approximation should be $\sim 10\%$ on decay
constants, $\sim 3\%$ on the ratios \fBsofB\ and \fDsofD,
and $\sim 5\%$ on the ratios \fBofDs\ and \fBsofDs. (The latter
ratios --- see Ref.~[19]
for current values ---
may prove the best way to get at \fB\ and \fBs, especially 
if  a future Tau-Charm factory allows for precise experimental
determination of \fDs.~\cite{stocchi})

The quenching errors on the decay constants will 
be more difficult to control, but I
expect  a gradual improvement in the estimates of this error.
As discussed above, the current estimates are rather crude,
so I emphasize that ``improvement in the estimates''
does not necessarily mean ``reduction of the error!''
In the MILC collaboration,  we expect to have a first 
extrapolation of virtual quark
results to $a\to0$ within one or two years.

Calculations of the B parameters are less far along, and it is therefore
more difficult to predict  what the improvement will be over
the next few years.  I am hopeful that studies of the lattice-spacing
dependence and of virtual quark effects, as well as perhaps higher
order perturbative calculations and/or nonperturbative evaluations
of the renormalization constants, will result in errors on $B_{B_d}$ of
$\sim 5\%$ within the quenched approximation and another $\sim 5\%$
due to quenching.  For the ratio $B_{B_s}/B_{B_d}$, which is already
quite well determined, one can hope to reduce the continuum extrapolation
and quenching errors to $1\%$.

\section*{Acknowledgments}
I thank S.\ Hashimoto of JLQCD, J.\ Simone and S.\ Ryan of the FNAL 
collaboration, and T. Draper of the Kentucky group for communicating their results to me prior to publication.  
I am also greatful to S.\ Hashimoto, as well as my colleagues in the MILC
collaboration, for useful discussions. This work was supported in part by
the Department of Energy, under grant number DE-FG02-91ER40628.


\section*{References}
\begin{thebibliography}{99}

\bibitem{wittig}H.\ Wittig, Oxford Preprint: OUTP-97-20P, May, 1997,
Submitted to {\em Int.\ J.\ Mod.\ Phys.\ A} 
(hep-lat/9705034).

\bibitem{onogi} T.\ Onogi, review \edinburgh.

\bibitem{bdhs}C.\ Bernard, T.\ Draper, G.\ Hockney and A.\ Soni,
\Journal{\PRD}{38}{3540}{1988}.

\bibitem{eichten} E.\ Eichten, \seillac, 170.

\bibitem{elc91}C.\ Allton \et, \Journal{\NPB}{349}{598}{1991}.

\bibitem{pcw91}C.\ Alexandrou \et, \Journal{\PLB}{256}{60}{1991}.

\bibitem{ape94}C.\ Allton \et, \Journal{\NPB}{413}{461}{1994}.

\bibitem{variational}A.\ Duncan, \et, \amsterdam, 433 and 441;
T.\ Draper and C.\ McNeile, \dallas, 453.

\bibitem{fnal95}A.\ Duncan, \et, \Journal{\PRD}{51}{5101}{1995}.

\bibitem{pcw94}C.\ Alexandrou, \et, \Journal{\ZPC}{62}{659}{1994}.

\bibitem{nrqcd}A.\ Ali Khan, \et, hep-lat/9704008.

\bibitem{nrqcd-japan}K-I.\ Ishikawa, \et, hep-lat/9706008.

\bibitem{lepage}G.P.\ Lepage and B.A.\ Thacker, \seillac, 199.

\bibitem{ekm}A.\ El-Khadra, A.\ Kronfeld, and
P.\ Mackenzie, \Journal\PRD{55}{3933}{1997}.

\bibitem{bls}C.\ Bernard, J.\ Labrenz and A.\ Soni,
 \Journal\PRD{49}{2536}{1994}.

\bibitem{elc92}A.\ Abada, \et,
 \Journal\NPB{376}{172}{1992}.

\bibitem{ukqcd}R.\ Baxter, \et,
 \Journal\PRD{49}{1594}{1994}.

\bibitem{ape97}C.\ Allton, \et, 
\Journal{\PLB}{405}{133}{1997}. 

\bibitem{milc-b20}for the current status 
of the MILC computation see C.\ Bernard, \et, 
talk presented at the conference {\it b20: Twenty Beautiful Years
of Bottom Physics}, Illinois Institute of Technology, June 29-July 2,
1997, to be published, 
hep-ph/9709328.  See also Ref.~[20].

\bibitem{milc-older} C.\ Bernard, \et,
\tsukubanew,
(hep-lat/9707013, July 1997); 
\stlouis, 358; \melbourne, 459; \bielefeld, 388.

\bibitem{jlqcd}S.\ Hashimoto, private communication, and
Ref.~[2].

\bibitem{fnal97}J.\ Simone, private communication, and
Ref.~[2].

\bibitem{clover}B.\ Sheikholeslami and R.\ Wohlert,
\Journal{\NPB}{259}{572}{1985}; G.\ Heatlie, \et,
\Journal{\NPB}{352}{266}{1991}.

\bibitem{qchpt} S.\ Sharpe, \prd{41}, 3233 (1990); C.\ Bernard and
M.\ Golterman, \prd{46}, 853 (1992).



\bibitem{soni} A.\ Soni, \melbourne, 43

\bibitem{recentBB} S.\ Aoki, \melbourne, 433.

\bibitem{oldBB}M.\ Gavela, \et, 
\Journal{\PLB}{206}{113}{1988}.

\bibitem{ukqcd96}A.\ Ewing, \et, 
\Journal{\PRD}{54}{3526}{1996}.

\bibitem{kentucky}  J.\  Christensen, T.\  Draper and C.\ McNeile
hep-lat/9610026.

\bibitem{gimenez} V.\ Gimenez and G.\ Martinelli,
\Journal{\PLB}{398}{135}{1997}.

\bibitem{blum}C.\ Bernard, T.\ Blum, and A.\ Soni, in preparation
and \stlouis, 382.

\bibitem{stocchi} P.\ Paganini, F.\ Parodi, P.\ Roudeau, and A.\ Stocchi,
``Measurements of the $\rho$ and $\eta$ parameters of the $V_{CKM}$
matrix and perspectives,'' submitted to CERN-PPE and {\em Phys.\ Rev.}\ D.



\end{thebibliography}
\end{document}